\documentclass[%
reprint,
superscriptaddress,
twocolumn,
amsmath,
amssymb,
aps,
prb
]{revtex4-1}

\usepackage{graphicx}  
\usepackage{dcolumn}  
\usepackage{bm}            
\usepackage{bbold}
\usepackage{amsmath}
\usepackage{xfrac}
\usepackage{xcolor}
\usepackage[colorlinks=true, citecolor=red]{hyperref}   
\usepackage{hhline}
\usepackage{diagbox}
\usepackage{siunitx}
\usepackage{enumitem}
\usepackage[normalem]{ulem}

\usepackage{siunitx}

\begin{document}
	
	\title{On the zero-field quantization of the  anomalous quantum Hall effect in moir\'e 2D layers}

	\author{Sankar Das Sarma}
	  	    \affiliation{Condensed Matter Theory Center and Joint Quantum Institute, Department of Physics, 
			                University of Maryland, College Park, Maryland 20742, USA}

	\author{Ming Xie}
    	\affiliation{Condensed Matter Theory Center and Joint Quantum Institute, Department of Physics, 
    		University of Maryland, College Park, Maryland 20742, USA}
    	                
	\date{\today}
	
	\begin{abstract}
		
		In  recent breakthrough experiments, twisted moir\'e layers of transition metal dichalcogenides are found to manifest both integer (IQAHE) and fractional (FQAHE) quantum anomalous Hall effects in zero applied magnetic field because of the underlying flat band topology and spontaneous breaking of the time reversal invariance.  In the current work, we critically analyze the experimental values of the quantized conductance in each case to emphasize the role of disorder in the problem, pointing out that obtaining accurate quantized conductance in future experiments would necessitate better contacts and lower disorder.

	\end{abstract}
	
	\maketitle

Recent experimental breakthroughs in multilayer moir\'e transition metal dichalcogenides (mTMD) 
\cite{Cai2023,Zeng2023, Park2023,Fan2023} 
as well as in pentalayer graphene \cite{PentaGraphene2023}
report the unambiguous observation of integer and fractional quantization of anomalous Hall effect without any Landau quantization imposed by an external magnetic field.  Although the effect was in some loose sense predicted theoretically \cite{Tang2011, Sun2011, Neupert2011, Regnault2011, Sheng2011, Shuo2012a, Shuo2012b}
in hypothetical flatband topological systems, nevertheless the remarkable discovery of the IQAHE and FQAHE in relatively dirty two-dimensional (2D) systems (with mobility $\sim$ 1000\,cm$^2$/Vs for the mTMDs)  with rather accurate, but not precise, resistance quantization has come as a huge surprise, and created considerable interest\cite{Li2021, ValentinFCI, ReddyFCI, ChongFCI, NicolasFCI, ParkerCFL, FuCFL, Yu2023}.

While much of the interest is on the origin of the interaction-induced zero-field fractional states, we focus on the issue of the accuracy of the quantization itself because of the extremely dirty nature of the underlying 2D materials. Why is the quantization seen at all in systems dominated by disorder?  
Most of the current and past theoretical literature on IQAHE and FQAHE completely ignore disorder 
and consider the emergence of topological quantum Hall states in clean systems. 
An earlier work by one of the authors takes into account its influence in a numerical exact diagonalization study
of a topological flatband model with onsite impurity potential \cite{Shuo2012a}.
In the current work, we quantitatively analyze the quantized resistance reported in the mTMD systems of 
Refs.~\onlinecite{Park2023, Fan2023} inferring that the expected future progress in sample fabrication and materials development should lead to (1) extremely accurate IQAHE quantization, possibly leading to a zero field resistance standard, and (2) observations of many more FQAHE states currently inhibited by disorder effects in the samples.

As a background, a mobility of  $\sim$1000 cm$^2$/Vs in samples of Refs.~\onlinecite{Park2023}
in twisted MoTe$_2$ homobilayers  \cite{Xuprivate} corresponds to a disorder induced broadening of $\sim\,$\SI{0.75}{\meV} ($\sim\,$\SI{9}{\kelvin}).
By contrast, the first observation  in 1982 of  the fractional quantum Hall effect (FQHE) at $1/3$ filling (and an applied field of \SI{15}{\tesla}) used a 2D GaAs sample with a mobility of 100,000 cm$^2$/Vs which corresponds to a disorder broadening of 
0.1 meV ($\sim\,$\SI{1}{\kelvin})---{the estimated FQHE gap was roughly \SI{1.5}{\kelvin}  in this sample} \cite{Tsui1982}.
It is well-known that FQHE is strongly adversely affected by disorder 
\cite{dAmbrumenil2011, Morf2002, Morfarxiv, Qian2017,Kleinbaum2020, Dean2008, Boebinger1987}.
Typically, the pristine fractional quantum Hall gap should be larger than the disorder strength for the fractional phase to be observable, implying that not only are all the currently observed FQAHE states in Refs.~\onlinecite{Park2023,Fan2023} likely to have much larger intrinsic gaps than the reported gaps in the existing dirty samples with an $\sim\,$\SI{9}{\kelvin} disorder, but also many more FQAHE phases with pristine gaps $<\,$\SI{9}{\kelvin} would manifest in future experiments as the system disorder is suppressed through materials improvement.  In particular, it is possible that FQAHE corresponding to non-Abelian fractional quantum Hall states at even denominator fillings would soon be observed in zero magnetic fields as system disorder improves since the connection between improved sample quality and robust even denominator FQHE is well-established \cite{Ahn2022,Chung2021}. 

By contrast, the integer quantum Hall effect (IQHE), being a single-particle topological effect directly manifesting a Chern number, should be stable to disorder unless the disorder is much larger than the (typically large) excitation gap for the IQHE (whence strong overlap among the disorder broadened Landau levels considerably complicate the physics).  In fact, disorder is essential in producing the quantized Hall resistance plateaus as it enables the chemical potential to move smoothly through the gapped localized regions between Landau levels---too much disorder, much larger than the pristine gap, complicates the situation by producing overwhelming localization, but with no disorder, there would be no quantum Hall plateaus whatsoever. In the recent mTMD QAHE experiments\cite{Cai2023,Park2023,Fan2023}, the zero-field integer excitation gap is estimated to be $>$40K, much larger than the disorder strength of $\sim$9K, so disorder should be irrelevant to the observed IQAHE.  The central paradigm of the quantum Hall effect  as arising from broadened Landau levels, where the disorder-induced localization at the edges of the Landau levels gives rise to the quantized plateaus in the transverse resistance and vanishing resistance in the longitudinal resistance, should apply to IQAHE as well.  At finite temperatures, the quantization of the Hall resistance (and the vanishing of the longitudinal resistance) is not perfect because of carrier activation and variable range hopping transport in the localized regime.  

We mention that the original discovery of IQHE already achieved a Hall resistance quantization close to 1 part in $10^5$ at \SI{1.8}{\kelvin} with the longitudinal resistance being essentially zero ($\sim$\SI{1}{\ohm}) \cite{Klitzing1980}
to be contrasted with the currently measured longitudinal resistance $>\,$\SI{100}{\ohm} in the AQHE experiment.  We believe that this difference arises entirely from the contact resistance problems in mTMD layers, where the noise arising from the contact resistance makes an accurate quantization a challenge independent of the cleanliness of the sample.  Future improvement in contact fabrication would be the key to progress in studying IQAHE and FQAHE.

\begin{figure}[b!]
	\includegraphics[width=0.5\textwidth]{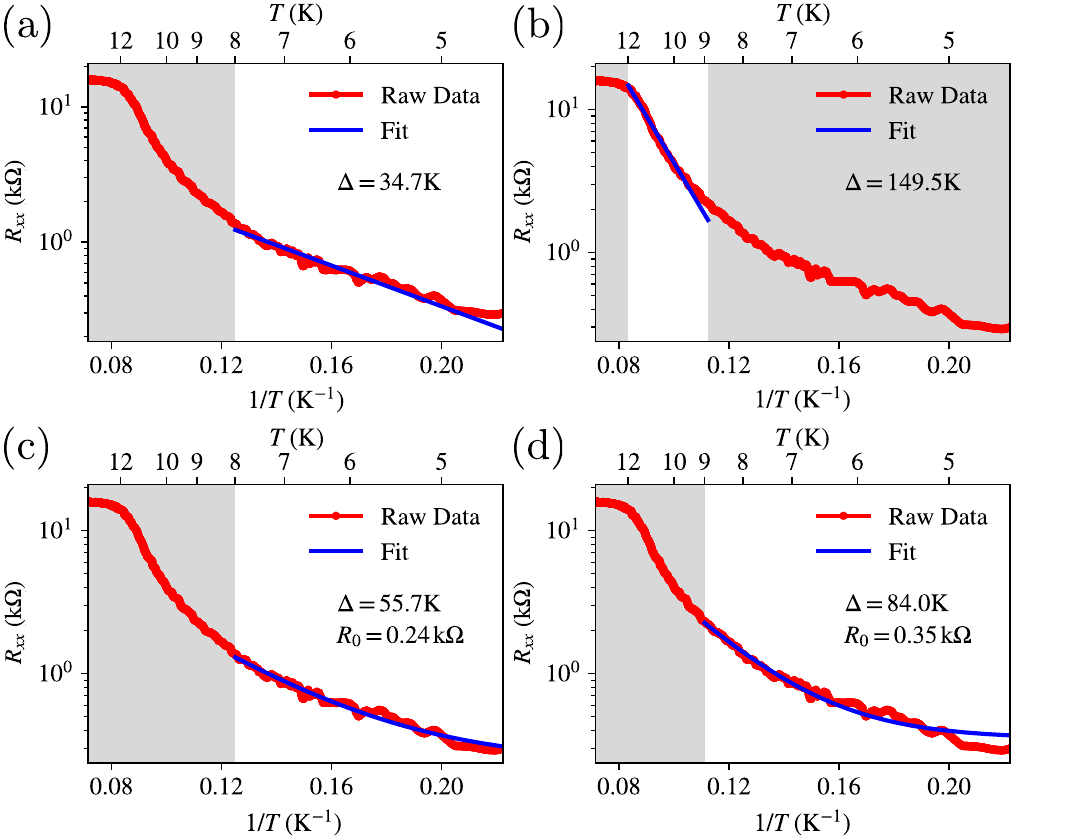}
	\caption{\label{activation} Thermal activation fitting of the longitudinal resistance $R_{xx}$ at filling factor $\nu=-1$
		and electric field $D/\epsilon_0=0$. 
		Data is for device $D(3.9^{\circ})$ taken from the Source Data of Fig.2d in Ref.~\onlinecite{Park2023}. 
		(a,b) are fittings without contact resistance and (c,d) are fittings with a contact resistance $R_0$ as a fitting parameter. The red lines are experiment data and blue lines
		are the fitting curves. Data in the shaded region is excluded from fitting.
	}
\end{figure}

In the rest of this work, we provide detailed quantitative fits to the experimental IQAHE and FQAHE finite temperature data to understand the quantization accuracy and the role of disorder.  In carrying out the fitting, it becomes clear that the contact resistance  plays a significant role as is obvious from the large background resistance as well the large amount of noise in the measured resistance even at cryogenic temperatures.  In fact, the most dramatic qualitative difference between the observed zero-field IQAHE/FQAHE  and the well-established high-field 2D IQHE/FQHE is that the large residual values of the longitudinal resistivity in the IQAHE/FQAHE  in the quantized Hall plateaus which tend to be vanishingly small in regular IQHE/FQHE.  
{We show through our quantitative analysis that this large residual resistance arises mainly from the uncontrolled contact problem in the current IQAHE/FQAHE samples.}
The exact nature of the residual resistance remains unclear and demands future experimental study.
We emphasize that we use the actual experimental data for our theoretical analysis with the novelty of our work being an unbiased analysis of the data in order to obtain objective quantitative conclusions about the experiments.

Our fittings use activation, Mott variable range hopping, and Efros-Shklovskii (ES) variable range hopping behaviors separately since which of these localized transport mechanisms is more dominant in a particular temperature range in the IQAHE/FQAHE measurements of Refs.~\onlinecite{Park2023,Fan2023} are a priori unknown.
In addition to the activated temperature dependence, we also carry out fitting including a contact resistance $R_0$ as a free fitting parameter,
\begin{figure}[t!]
	\includegraphics[width=0.5\textwidth]{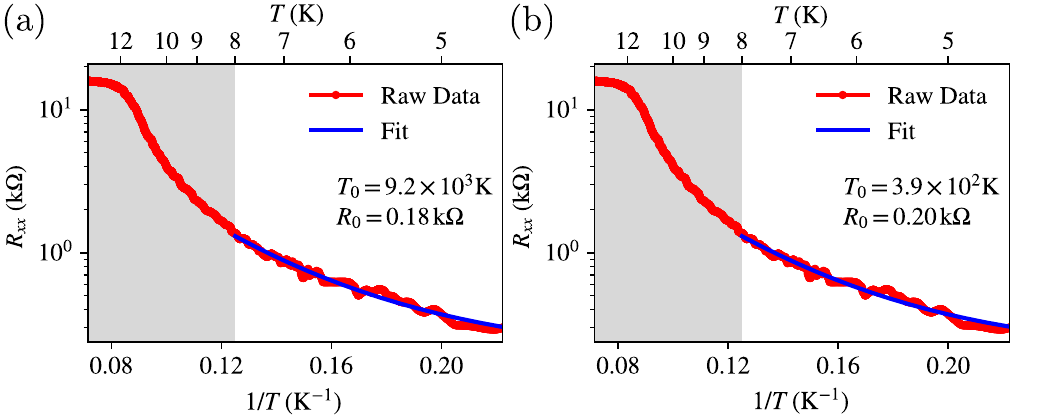}
	\caption{\label{VRfittings} 
		(a) Mott and (b) Efros-Shklovskii variable range hopping fitting of the longitudinal resistance $R_{xx}$ at filling factor $\nu=-1$ and electric field $D/\epsilon_0=0$. 
		The source data is the same as used in Fig.~\ref{activation}. 
		The red lines are experiment data and blue lines
		are the fitting curves. Data in the shaded region is excluded from fitting.
		$T_0$ is a parameter in variable range hopping models in Eq.~(\ref{mott}) and (\ref{ES}).
}
\end{figure}
\begin{figure}[b!]
	\includegraphics[width=.5\textwidth]{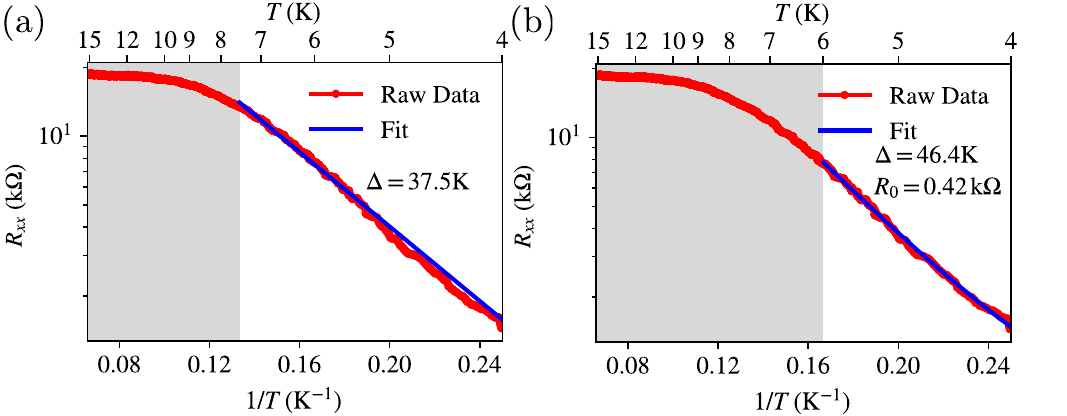}
	\caption{\label{activation35} Thermal activation fitting of the longitudinal resistance $R_{xx}$ at filling factor $\nu=-1$
		and electric field $D/\epsilon_0=0$ for device $D(3.5^{\circ})$.
Data is taken from the Source Data of Extended Fig.2b in Ref.~\onlinecite{Park2023}. 
			(a) and (b) are fittings without and with a contact resistance $R_0$ as a fitting parameter, respectively. The red lines are experiment data and blue lines are the fitting curves. Data in the shaded regions is excluded from fitting.
	}
\end{figure}
\begin{figure*}
	\includegraphics[width=.8\textwidth]{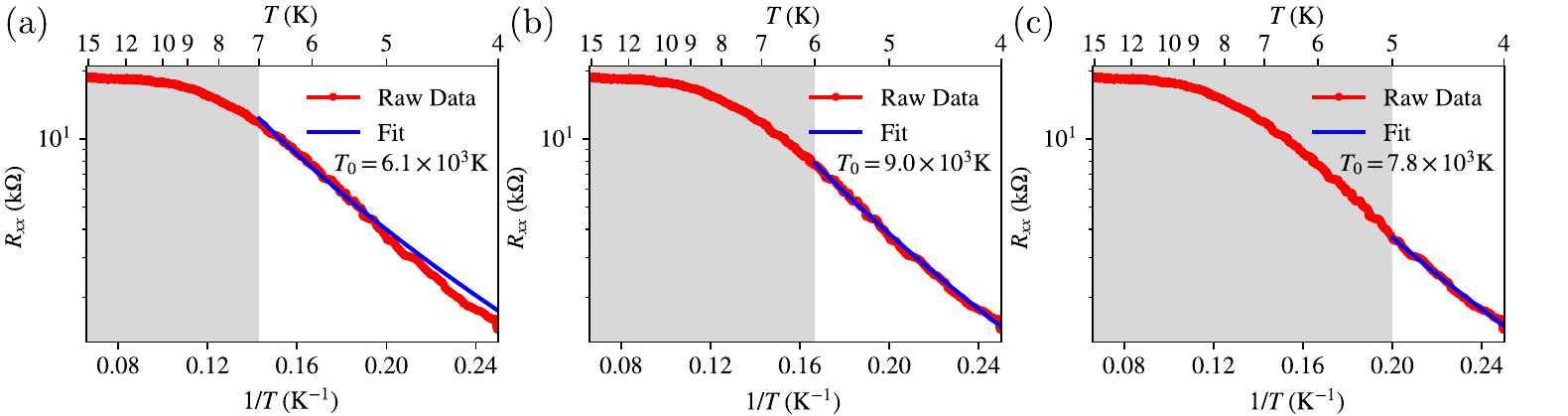}
	\caption{\label{mott35} Mott variable range fitting of the same data as presented in Fig.~\ref{activation35}. 
         From (a) to (c), the fitting range in $T^{-1}$ is decreased.	
         $R_0$ is set to zero in all three cases.
	}
\end{figure*}
\begin{align}
	R(T) & = R_0 + Ae^{-\Delta/2k_BT} 
	\label{activiation}
\end{align}
where $\Delta$ is the thermal activation gap and $A$ is the proportional constant which is also a fitting parameter.
In the presence of disorder, variable range hopping mechanisms can also manifest, thereby contributing to the electrical resistance observed.
In two-dimensional systems, the temperature dependence of the resistance takes the form
\begin{align}
		R(T) &= R_0 + Ae^{-(T_0/T)^{1/3}}
		\label{mott}
\end{align}
for Mott variable range hopping \cite{Mott1969} and
\begin{align}
	R(T) &= R_0 + Ae^{-(T_0/T)^{1/2}}
	\label{ES}
\end{align}
for ES variable range hopping\cite{ES1975}.
In both cases, we carry out fit with both $R_0$ set to 0 and $R_0$ as a free fitting parameter.
$A$ and $T_0$ are model dependent constants treated as free fitting parameters as well.

We analyze two situations from Ref.~\onlinecite{Park2023} for the IQAHE: two different devices $D(3.9^{\circ})$ and $D(3.5^{\circ})$ with twist angle $\theta=3.9^{\circ}$ and $3.5^{\circ}$ respectively.  
We also analyze the IQAHE data of Ref.~\onlinecite{Fan2023}.
{\color{red} Figure~\ref{activation}} show the thermal activation fits for $\theta=3.9^{\circ}$ with the top two panels without any contact resistance, and the bottom two with a contact resistance $R_0$ which is extracted from the best fits.
The top two panels show that the fits provide very different activation energies depending on the fit regions, whereas the bottom two panels, including the contact resistance, achieve decent fits over the whole regime with mutually consistent residual (contact) resistance and energy gaps.  
If we restrict our activation fits to lower temperatures, where activation typically dominates, we get $R_0\sim 300 ~\Omega$ and gap $\sim$ 70 K.  We note that the extracted $R_0\sim 300 ~\Omega$ is $2-3$ orders of magnitude larger than in IQHE for 2D semiconductors and graphene, emphasizing the considerable scope for materials improvement in IQAHE samples.  
The two panels of {\color{red} Fig.~\ref{VRfittings}} show the Mott and ES variable range hopping fits including the contact resistance, and the fits are both reasonable, but the important point is that the extracted contact resistance is roughly consistent with that in Fig.~\ref{activation}.
Currently existing experimental results cannot sharply distinguish among the localized transport mechanisms, but they all lead to similar contact resistance which is much larger than the corresponding residual resistance in regular IQHE.

 In  {\color{red}Fig.~\ref{activation35}}, we show the activation fits to the longitudinal resistance at $\theta=3.5^{\circ}$, finding again a gap $\sim$ 40K, 
However, the fitted contact resistance can be much larger in this case $R_0\sim0.4\,-$\SI{1.5}{\kohm}, depending on the choice of cutoff,
 which cannot really be precisely pinned down given the limitation of the available experimental data.  
 In {\color{red}Fig.~\ref{mott35}}, we reinforce the inadequacy of the current data in precisely pinning down the localized transport mechanism by showing fits to Mott variable range hopping by modifying the regime of fitting---all give good fits ($R_0$=0 in these fits to reduce the number of unknowns), but the fitting parameters are different.  
 Similar behavior is also seen in the ES fits (not shown).

\begin{figure*}
	\includegraphics[width=0.8\textwidth]{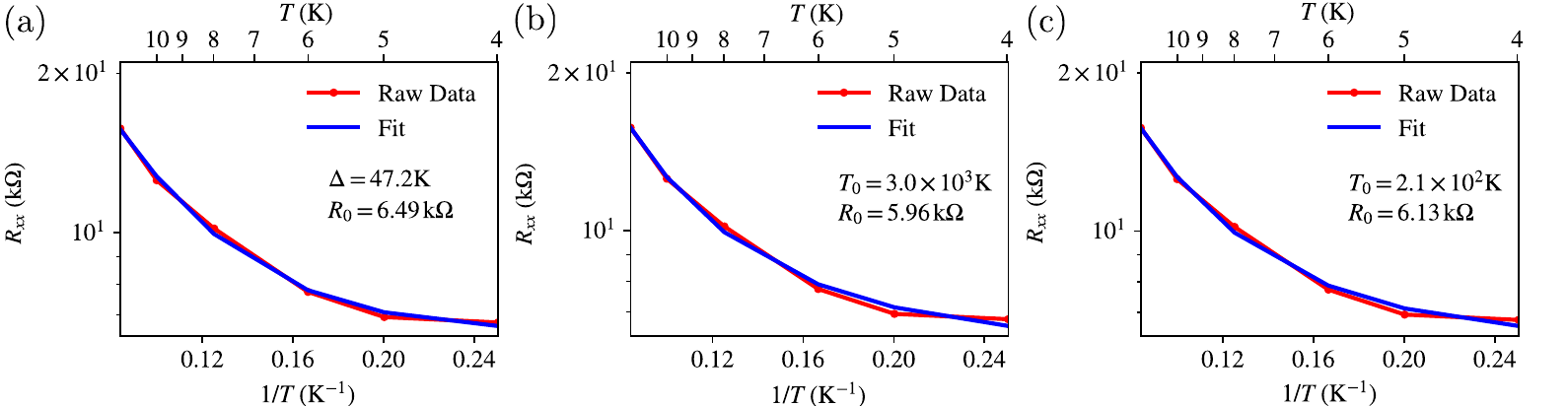}
	\caption{\label{PRX} (a) Thermal activation, (b) Mott variable range hopping, and (c) ES variable range hopping fittings of the longitudinal resistance $R_{xx}$ at filling factor $\nu=-1$ in Ref.~\onlinecite{Fan2023}.  The red lines are experiment data and blue lines are the fitting curves. 
		The data is extracted from SFig.4b for $D=-15\,$mV/nm in Ref.~\onlinecite{Fan2023}.
	}
\end{figure*}

In {\color{red} Fig.~\ref{PRX}}, we show resistance fits for the IQAHE data from the Ref.~\onlinecite{Fan2023} using activation, Mott, and ES models including a background resistance $R_0$ in all cases. 
The fitted gap is around $\sim$\SI{47}{\kelvin} for the thermal activation with contact resistance, 
which is 2-3 times of the value without considering the contact resistance ($\sim10-$\SI{20}{\kelvin} according to Fig.3c of Ref.~\onlinecite{Fan2023}).
The contact resistances $R_0\sim5-$\SI{7}{\kohm} are similar and significant from all the fitting scenarios, showing that the physics is likely to be dominated by the contact resistance problem.  We emphasize that similar analyses give vanishing $R_0$ ($\sim$\SI{1}{\ohm}) for regular high field 2D IQHE phenomena, reiterating the serious sample problem plaguing the zero field IQAHE/FQAHE physics. 
The contact resistance in the sample of Ref.~\onlinecite{Fan2023} is an order of magnitude larger
than that of the devices in Ref.~\onlinecite{Park2023}.
Our work establishes that the difference in the degree of quantization between them is most likely not arising from the intrinsic IQAHE excitation gap ($\sim$40K in both experiments), but can be improved through materials improvement.

\begin{figure}[b!]
	\includegraphics[width=.49\textwidth]{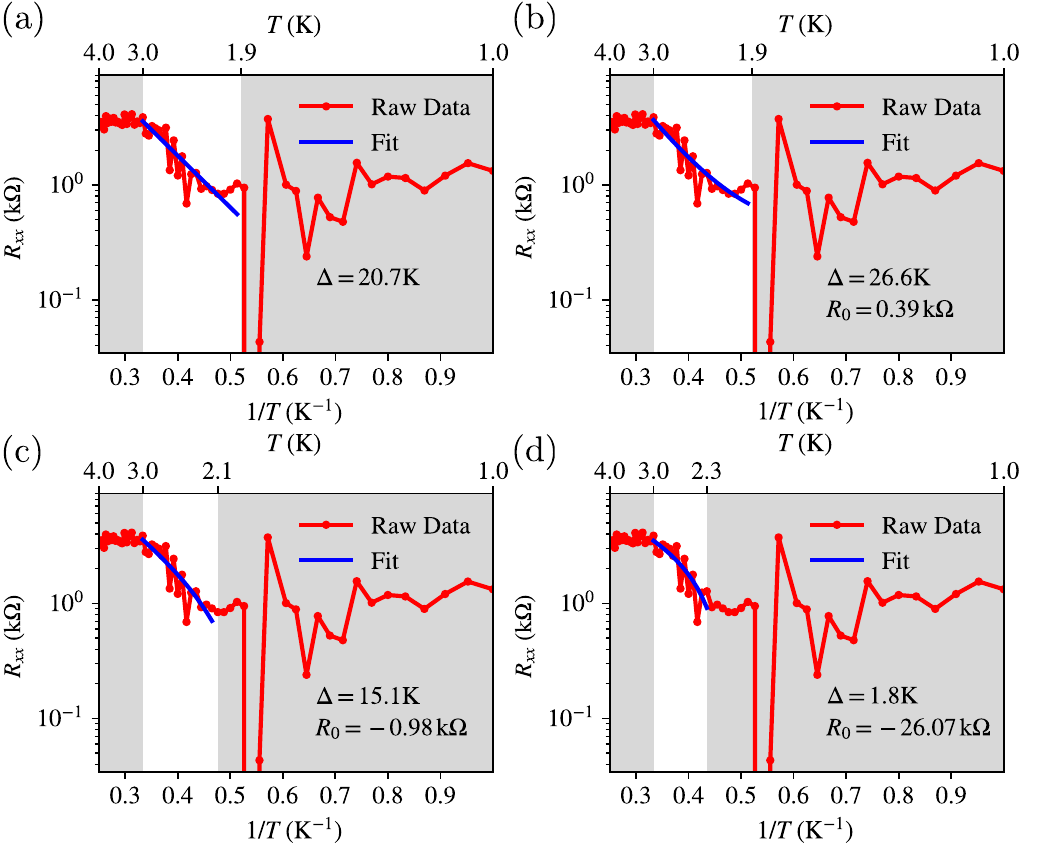}
	\caption{\label{2thirds} 
		Thermal activation fitting of the $R_{xx}$ at filling factor $\nu=-2/3$
		and electric field $D/\epsilon_0=0$.
Data is for device $D(3.7^{\circ})$ taken from the Source Data of Extended Fig.5a in Ref.~\onlinecite{Park2023}. 
		(a) is fittings without contact resistance and (b-d) are fittings with a contact resistance $R_0$. The red lines are experiment data and blue lines
		are the fitting curves. Data in the shaded region is excluded from fitting.
		Note the very small regimes of the data where any fitting is meaningful.
	}
\end{figure}

In the Supplemental Material \cite{supplement}, we provide detailed results on the deviations of the measured longitudinal resistance from zero and the Hall resistance from the quantized value (as well as an empirical relationship between them \cite{ResistanceStandard})  as a function of temperature, displacement field to clearly demonstrate that the IQAHE quantization is quite inaccurate and is very far from being useful for a resistance standard.  But sample improvements should make the situation better in the future. 
\begin{figure*}
	\includegraphics[width=.8\textwidth]{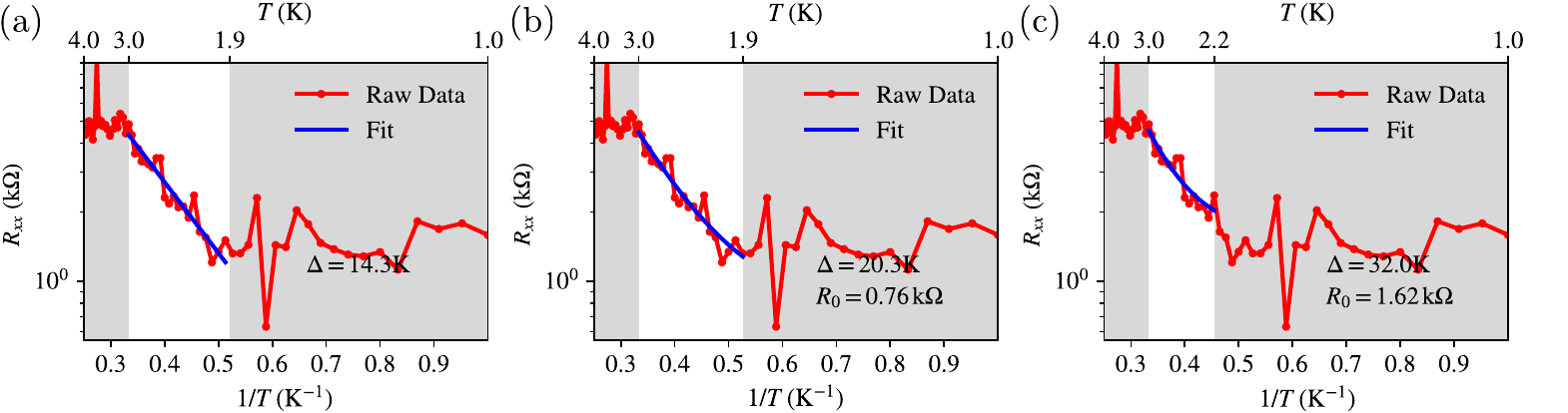}
	\caption{\label{3fifths} Thermal activation fitting of the $R_{xx}$ at filling factor $\nu=-3/5$
		and electric field $D/\epsilon_0=0$.	
Data is for device $D(3.7^{\circ})$ taken from the Source Data of Extended Fig.6b in Ref.~\onlinecite{Park2023}. 
(a) is fittings without contact resistance and (b,c) are fittings with a contact resistance $R_0$.  
		The red lines are experiment data and blue lines are the fitting curves. Data in the shaded region is excluded from fitting.
	}
\end{figure*}


Now, we discuss, for the sake of completeness,  the corresponding analyses of the resistivity data for FQAHE, which, we must warn, is hopelessly inadequate because of the huge fluctuations in the measured resistance.  
In {\color{red}Fig.~\ref{2thirds}}, we show the activation fits to the resistivity for $\theta=3.7^{\circ}$ and the fractional filling of $-2/3$. 
The four panels show fits to the same data using $R_0=0$ and finite $R_0$ varying the fit regimes.  The extracted FQAHE gap is $\sim$\SI{23}{\kelvin} without any contact resistance correction, but with a finite $R_0$ the situation becomes intractable because of the paucity of data with both the gap and $R_0$ changing with varying fit regimes.  We emphasize that the available data are so noisy that any fit works only over a very narrow range of temperature ($\sim$ 1K only!), and not much significance should be attached to these fits.  
The fact that in some fitting windows $R_0$ even becomes negative may suggest that the transport
behavior cannot be well captured by the thermal activation model if we were to rely on the limited data.
The only conclusion is that an activation gap $\sim$\SI{23}{\kelvin} is measured over a narrow regime, but the quantization itself is extremely noisy, most likely because of considerable measurement noise arising from the contact resistance.  In {\color{red} Fig.~\ref{3fifths}}, we show similar activation fits to the $-3/5$ FQAHE in Ref.~\onlinecite{Park2023} for $\theta=3.7^{\circ}$, finding again that the resistivity is  simply too noisy for any decisive conclusion.  The gap at $-3/5$ varies between \SI{14}{\kelvin} and \SI{32}{\kelvin} depending on the details of the fits, all of which are, however, over very narrow regimes of temperatures.  The extracted contact resistance is large, $R_0\sim0.7-$\SI{1.6}{\kohm}, again showing that the contact resistance problem is severe in the IQAHE/FQAHE measurements.  The corresponding regular high-field FQHE measurements basically manifest zero contact resistance, and the activation fits work over at least one decade in temperature.  It is clear that the noise problem arising from contact resistance is huge in the current IQAHE/FQAHE experiments, and any firm quantitative conclusions on the quantization accuracy must await much better samples with much less noise and much smaller contact resistance.  Our similar analysis (not shown) for the FQAHE data at filling $-2/3$ in Ref.~\onlinecite{Fan2023} has similar fitting problems, and the extracted contact resistance $R_0>$\SI{1}{\kohm} (being consistent with the corresponding IQAHE analysis) with the extracted gap being very small ($\sim$\SI{2}{\kelvin}), indicating a rather disordered sample.

\begin{figure}
	\includegraphics[width=.49\textwidth]{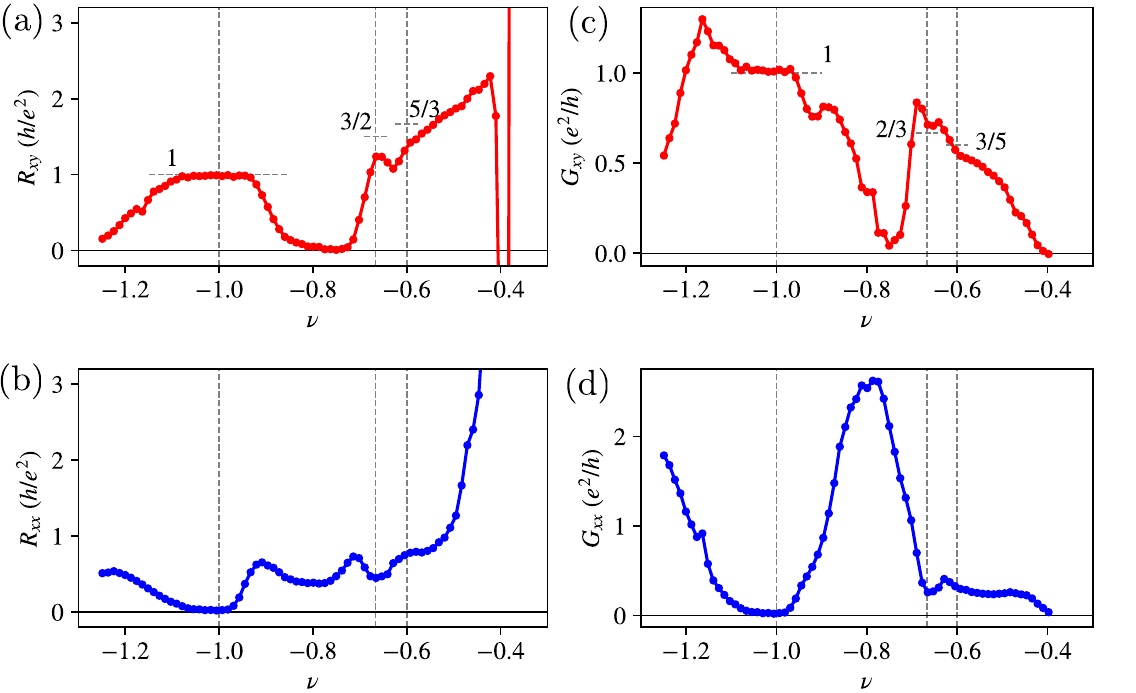}
	\caption{\label{Rvsnuplot} Filling factor dependence of the Hall (a) and longitudinal (b) resistances, and the corresponding Hall (c) and longitudinal (d) conductances. Vertical dashed lines mark filling factor $\nu=-1,-2/3$ and $-3/5$ and the horizontal dashed line segments mark the expected integer and fractional quantization values for $R_{xy}$ in (a) and $G_{xy}$ in (c). Data is taken from the Source Data of Fig.1(c) and (d) in Ref.~\onlinecite{Park2023} near zero displacement field $D/\epsilon_0=-0.12$ mV$\cdot$nm$^{-1}$. }
\end{figure}

Finally, in Fig.~\ref{Rvsnuplot}, we provide the resistances and corresponding conductances calculated from 
$G_{\alpha\beta}=R_{\alpha\beta}/(R_{xx}^2+R_{xy}^2)$, with $\alpha,\beta=x,y$,
as a function of filling factor, based on the experimental data of Ref.~\onlinecite{Park2023}.
The filling factor dependent results in Fig.~\ref{Rvsnuplot} decisively establish that although the current experiments provide a reasonable quantization for the IQAHE, the corresponding quantization for FQAHE is less accurate because of disorder effects.  This is, of course, similar to the situation arising historically in the original IQHE \cite{Klitzing1980} and FQHE \cite{Tsui1982} measurements, where fractional quantization turns out to be much more sensitive to disorder than the integer quantization.
While the Hall conductance appears to be more precisely quantized compared to the Hall resistance, it is the Hall resistance that is directly measured when a current source is supplied (instead of a bias voltage), and therefore, it more accurately reflects the true precision of the quantization.

In conclusion, we have carried out a detailed quantitative analysis of the recently reported IQAHE and FQAHE resistivity results, finding a large contact resistance, considerably complicating the physics of quantization even when the extracted excitation gap is large.  
Unlike the IQHE and FQHE, where the gap sizes are governed by two different energy scales—the cyclotron energy and the interaction energy—the gaps in both the IQAHE and FQAHE are a consequence of strong electron interaction. This leads to gap sizes in both IQAHE and FQAHE depending on the interaction energy scale and appearing to be of a similar order, as observed in experiments.
We remark that four probe transport measurements should eliminate the usual contact resistance artifacts, and therefore our finding of a large residual resistance (of unknown origin) through a direct analysis of the experimental data is somewhat of a mystery.
In addition, the contact resistance is very noisy, producing large resistance fluctuations, which compromise the quantization accuracy.  Further materials improvement and better sample fabrication  should improve the situation, but until that happens, the zero field IQAHE/FQAHE will not be useful in the implementation of zero field resistance standards.

\begin{acknowledgments}
	{\em Acknowledgment.}---\noindent	The authors acknowledge helpful discussions with Xiaodong Xu, Jiaqi Cai,  Tingxin Li and Long Ju.  This work is supported by Laboratory for Physical Sciences.
\end{acknowledgments}

\begin{thebibliography}{99}
	
\bibitem{Cai2023}
J. Cai, E. Anderson, C. Wang, X. Zhang, X. Liu, W. Holtzmann, Y. Zhang, F. Fan, T. Taniguchi, K. Watanabe, Y. Ran, T. Cao, L. Fu, D. Xiao, W. Yao, and X. Xu,
Signatures of fractional quantum anomalous Hall states in twisted MoTe$_2$,
Nature (London) \textbf{622}, 63 (2023).

\bibitem{Zeng2023}
Y. Zeng, Z. Xia, K. Kang, J. Zhu, P. Knüppel, C. Vaswani, K. Watanabe, T. Taniguchi, K. Fai Mak, and J. Shan, 
Thermodynamic evidence of fractional Chern insulator in moiré MoTe2
Nature (London) \textbf{622}, 69 (2023)

\bibitem{Park2023}
H. Park, J. Cai, E. Anderson, Y. Zhang, J. Zhu, X. Liu, C. Wang, W. Holtzmann, C. Hu, Z. Liu, T. Taniguchi, K. Watanabe, J.-H. Chu, T. Cao, L. Fu, W. Yao, C.-Z. Chang, D. Cobden, D. Xiao, and X. Xu,
Nature  (London) \textbf{622}, 74 (2023).

\bibitem{Fan2023}
F. Xu, Z. Sun, T. Jia, C. Liu, C. Xu, C. Li, Y. Gu, K. Watanabe, T. Taniguchi, B. Tong, J. Jia, Z. Shi, S. Jiang, Y. Zhang, X. Liu, and T. Li,
Observation of Integer and Fractional Quantum Anomalous Hall Effects in Twisted Bilayer MoTe$_2$,
Phys. Rev. X \textbf{13}, 031037 (2023).

\bibitem{PentaGraphene2023}
Z. Lu, T. Han, Y. Yao, A. P. Reddy, J. Yang, J. Seo, K. Watanabe, T. Taniguchi, L. Fu, and L. Ju,
Fractional Quantum Anomalous Hall Effect in a Graphene Moire Superlattice,
arXiv:2309.17436.


\bibitem{Tang2011}
E. Tang, J.-W. Mei, and X.-G. Wen,
High-Temperature Fractional Quantum Hall States,
Phys. Rev. Lett. \textbf{106}, 236802 (2011).

\bibitem{Sun2011}
K. Sun, Z. Gu, H. Katsura, and S. Das Sarm,
Nearly Flatbands with Nontrivial Topology,
Phys. Rev. Lett. 106, 236803 (2011).

\bibitem{Neupert2011}
T. Neupert, L. Santos, C. Chamon, and C. Mudry,
Fractional Quantum Hall States at Zero Magnetic Field,
Phys. Rev. Lett. \textbf{106}, 236804 (2011).


\bibitem{Regnault2011}
N. Regnault and B. A. Bernevig,
Fractional Chern Insulator,
Phys. Rev. X \textbf{1}, 021014 (2011).

\bibitem{Sheng2011}
D. N. Sheng, Z.-C. Gu, K. Sun, and L. Sheng,
Fractional quantum Hall effect in the absence of Landau levels,
Nat. Comm. \textbf{2}, 389 (2011).

\bibitem{Shuo2012a}
S. Yang, K. Sun, and S. Das Sarma,
Quantum phases of disordered flatband lattice fractional quantum Hall systems,
Phys. Rev. B \textbf{85}, 205124 (2012).

\bibitem{Shuo2012b}
S. Yang, Z.-C. Gu, K. Sun, and S. Das Sarma,
Topological flat band models with arbitrary Chern numbers,
Phys. Rev. B \textbf{86}, 241112(R) (2012).




\bibitem{Li2021}
H. Li, U. Kumar, K. Sun, and S.-Z. Lin,
Spontaneous fractional Chern insulators in transition metal dichalcogenide moir\'e superlattices,
Phys. Rev. Research \textbf{3}, L032070 (2021).

\bibitem{ValentinFCI}
V. Crépel and L. Fu,
Anomalous Hall metal and fractional Chern insulator in twisted transition metal dichalcogenides,
Phys. Rev. B \textbf{107}, L201109 (2023).

\bibitem{ReddyFCI}
A. P. Reddy, F. Alsallom, Y. Zhang, T. Devakul, and L. Fu,
Fractional quantum anomalous Hall states in twisted bilayer MoTe2 and WSe2,
Phys. Rev. B \textbf{108}, 085117 (2023).

\bibitem{ChongFCI}
C. Wang, X.-W. Zhang, X. Liu, Y. He, X. Xu, Y. Ran, T. Cao, and D. Xiao,
Fractional Chern Insulator in Twisted Bilayer MoTe2,
arXiv:2304.11864.
 

\bibitem{NicolasFCI}
N. Morales-Dur\'an, J. Wang, G. R. Schleder, M. Angeli, Z. Zhu, E. Kaxiras, C. Repellin,and J. Cano, 
Pressure–enhanced fractional Chern insulators in moir\'e transition metal dichalcogenides along a magic line,
arXiv:2304.06669.



\bibitem{ParkerCFL}
J. Dong, J. Wang, P. J. Ledwith, A. Vishwanath, and D. E. Parker,
Composite Fermi Liquid at Zero Magnetic Field in Twisted 
MoTe$_2$,
Phys. Rev. Lett. \textbf{131}, 136502 (2023).

\bibitem{FuCFL}
H. Goldman, A. P. Reddy, N. Paul, and L. Fu,
Zero-Field Composite Fermi Liquid in Twisted Semiconductor Bilayers,
Phys. Rev. Lett. 131, 136501 (2023).

\bibitem{Yu2023}
J. Yu, J. Herzog-Arbeitman, M. Wang, O. Vafek, B. A. Bernevig, and N. Regnault, 
Fractional Chern Insulators vs. Non-Magnetic States in Twisted Bilayer MoTe$_2$,
arXiv:2309.14429 (2023).

\bibitem{Xuprivate}
X. Xu, private communication, 2023.

\bibitem{Tsui1982}
D. C. Tsui, H. L. Stormer, and A. C. Gossard,
Two-Dimensional Magnetotransport in the Extreme Quantum Limit,
Phys. Rev. Lett. \textbf{48}, 1559 (1982).


\bibitem{dAmbrumenil2011}
N. d’Ambrumenil, B. I. Halperin, and R. H. Morf,
Model for Dissipative Conductance in Fractional Quantum Hall States,
Phys. Rev. Lett. \textbf{106}, 126804 (2011).

\bibitem{Morf2002}
R. H. Morf, N. d’Ambrumenil, and S. Das Sarma,
Excitation gaps in fractional quantum Hall states: An exact diagonalization study,
Phys. Rev. B \textbf{66}, 075408 (2002).

\bibitem{Morfarxiv}
R. H. Morf, N. d'Ambrumenil,
The Gap at nu = 5/2 and the Role of Disorder in Fractional Quantum Hall States,
arXiv:0212304.

\bibitem{Qian2017}
Q. Qian, J. Nakamura, S. Fallahi, G. C. Gardner, J. D. Watson, S. Lüscher, J. A. Folk, G. A. Csáthy, and M. J. Manfra,
Quantum lifetime in ultrahigh quality GaAs quantum wells: Relationship to $\Delta_{5/2}$ and impact of density fluctuations,
Phys. Rev. B \textbf{96}, 035309 (2017).

\bibitem{Kleinbaum2020}
E. Kleinbaum, Hongxi Li, N. Deng, G. C. Gardner, M. J. Manfra, and G. A. Csáthy,
Disorder broadening of even-denominator fractional quantum Hall states in the presence of a short-range alloy potential,
Phys. Rev. B \textbf{102}, 035140 (2020).

\bibitem{Dean2008}
C. R. Dean, B. A. Piot, P. Hayden, S. Das Sarma, G. Gervais, L. N. Pfeiffer, and K. W. West,
Intrinsic Gap of the $\nu$=5/2 Fractional Quantum Hall State,
Phys. Rev. Lett. \textbf{100}, 146803 (2008).

\bibitem{Boebinger1987}
G. S. Boebinger, H. L. Stormer, D. C. Tsui, A. M. Chang, J. C. M. Hwang, A. Y. Cho, C. W. Tu, and G. Weimann,
Activation energies and localization in the fractional quantum Hall effect,
Phys. Rev. B \textbf{36}, 7919 (1987).

\bibitem{Ahn2022}
S. Ahn and S. Das Sarma,
Density-dependent two-dimensional optimal mobility in ultra-high-quality semiconductor quantum wells,
Phys. Rev. Materials \textbf{6}, 014603 (2022).

\bibitem{Chung2021}
Y. J. Chung, K. A. Villegas Rosales, K. W. Baldwin, P. T. Madathil, K. W. West, M. Shayegan, and L. N. Pfeiffer,
Ultra-high-quality two-dimensional electron systems,
Nat. Mat. \textbf{20}, 632 (2021).

\bibitem{Klitzing1980}
K. v. Klitzing, G. Dorda, and M. Pepper,
New Method for High-Accuracy Determination of the Fine-Structure Constant Based on Quantized Hall Resistance,
Phys. Rev. Lett. \textbf{45}, 494 (1980).

\bibitem{Mott1969}
N. F. Mott,
Conduction in non-crystalline materials, Philosophical Magazine
\textbf{19}, 835 (1969).

\bibitem{ES1975}
A. L. Efros and B. I. Shklovskii, 
Coulomb Gap and Low Temperature Conductivity of Disordered Systems. 
J. Phys. C \textbf{8}, L49 (1975).

\bibitem{supplement}
See Supplemental Material at [url] for detailed results on the deviations of the measured longitudinal resistance from zero and the Hall resistance from the quantized value (as well as an empirical relationship between them)  as a function of temperature and displacement field. The Supplemental Material also contains Ref.~\onlinecite{ResistanceStandard}.

\bibitem{ResistanceStandard}
J. Matthews and M. E. Cage,
Temperature Dependence of the Hall and Longitudinal Resistances in a Quantum Hall Resistance Standard,
J. Res. Natl. Inst. Stand. Technol., \textbf{110}, 497 (2005).

\end{thebibliography}
\end{document}